\documentstyle[12pt]{article}
\normalbaselines

\begin{document}
\begin{flushright}

{\raggedleft

HUB-EP-97/83\\

hep-th/9712113\\[1cm]}

\end{flushright}

\renewcommand{\thefootnote}{\fnsymbol{footnote}}
\begin{center}
{\LARGE\baselineskip0.9cm 
A special irreducible matrix representation 

of the real Clifford algebra C(3,1)\\[1.5cm]}

{\large K. Scharnhorst\footnote[2]{E-mail:
{\tt scharnh@physik.hu-berlin.de}}
}\\[0.3cm]

{\small Humboldt-Universit\"at zu Berlin

Institut f\"ur Physik

Invalidenstr.\ 110

D-10115 Berlin

Federal Republic of Germany}\\[1.5cm]

\begin {abstract}
$4\times 4$ Dirac (gamma) matrices (irreducible matrix 
representations of the Clifford algebras C(3,1), C(1,3), C(4,0)) 
are an essential part of many calculations in quantum physics.
Although the final physical results do not depend on 
the applied representation of the Dirac matrices (e.g.\ due to the invariance
of traces of products of Dirac matrices), the appropriate choice of 
the representation used may facilitate the analysis. The present
paper introduces a particularly symmetric real representation of 
$4\times 4$ Dirac matrices (Majorana representation) which 
may prove useful in the future. As a byproduct, a compact 
formula for (transformed) Pauli matrices is found.
The consideration is based on the role 
played by isoclinic 2-planes in the geometry of the real 
Clifford algebra C(3,0) which provide an invariant geometric 
frame for it. It can be generalized to larger 
Clifford algebras.
\end{abstract}

\end{center}

\renewcommand{\thefootnote}{\arabic{footnote}}

\thispagestyle{empty}

\newpage
\section{Introduction}

Dirac (gamma) matrices used within many calculations in quantum physics
can be understood as representations of Clifford algebras.
In 4D Minkowski or Euclidean space they are representations of the 
Clifford algebras C(3,1), C(1,3) or C(4,0), respectively. 
While there is no problem to write down sets of complex $4\times 4$
Dirac matrices which form irreducible representations of these
Clifford algebras, a set of real $4\times 4$ Dirac matrices (Majorana
representation), which we will be interested in, 
can only be obtained for the Clifford algebra C(3,1)
\cite{okub1}-\cite{brih} (further material on real Clifford algebras
can be found in \cite{port}, ch.\ 13, 
\cite{port2}-\cite{shaw}). These matrices obey 
the standard relation
\begin{eqnarray}
\label{A1}
\gamma_\mu \gamma_\nu\ +\ \gamma_\nu \gamma_\mu &=& 
2\ \eta_{\mu\nu}\ {\bf 1}
\end{eqnarray}
where $\eta_{\mu\nu}$, $\mu,\nu = 1,..,4$ are the elements of 
the diagonal matrix $\eta$ with ${\rm diag}(\eta) = (1,1,1,-1)$
and ${\bf 1}$ is the $4\times 4$ unit matrix.
An explicit representation of real gamma matrices is provided 
by the following expressions (adapted from \cite{brih}).

\parbox{7cm}{
\begin{eqnarray}
\label{A2a}
\gamma_1&=&
\left(
\begin{array}{*{4}{c}}
0&0&0&1\\ 
0&0&1&0\\
0&1&0&0\\
1&0&0&0
\end{array}
\right)
\end{eqnarray}
}
\hfill
\parbox{7cm}{
\begin{eqnarray}
\label{A2b}
\gamma_2&=&
\left(
\begin{array}{*{4}{c}}
0&0&1&0\\ 
0&0&0&-1\\
1&0&0&0\\
0&-1&0&0
\end{array}
\right)
\end{eqnarray}
}

\parbox{7cm}{
\begin{eqnarray}
\label{A2c}
\gamma_3&=&
\left(
\begin{array}{*{4}{c}}
1&0&0&0\\ 
0&1&0&0\\
0&0&-1&0\\
0&0&0&-1
\end{array}
\right)
\end{eqnarray}
}
\hfill
\parbox{7cm}{
\begin{eqnarray}
\label{A2d}
\gamma_4&=&
\left(
\begin{array}{*{4}{c}}
0&0&1&0\\ 
0&0&0&1\\
-1&0&0&0\\
0&-1&0&0
\end{array}
\right)
\end{eqnarray}
}

\noindent
But, eq.\ (\ref{A1}) is invariant under orthogonal transformations
$O$ of the gamma matrices 
\begin{eqnarray}
\label{A3}
\gamma^\prime_\mu = O\gamma_\mu O^T 
\end{eqnarray}
and any other set of congruent (by virtue of (\ref{A3}))
gamma matrices $\gamma^\prime_\mu$ will also be equally 
appropriate as representation of C(3,1) (the general 
situation is described by Pauli's fundamental 
theorem \cite{paul}, \cite{good}). Now, let us denote 
the real linear vector space ${\bf R}_4$ in which the 
elements of the Clifford algebra C(3,1) act as operators 
by $V$ (spinor space). Then,
the matrices $\gamma_\mu$ can be understood as representations of
the generators of C(3,1) with respect to a certain orthonormal basis
in $V$ which defines in it a rectangular coordinate system.
Any transformation (\ref{A3}) of the gamma matrices corresponds
to an orthogonal transformation in $V$ and consequently to a change
of the coordinate system in $V$. The concrete shape of the gamma
matrices changes in performing these transformations. In 
explicit calculations in which gamma matrices occur the required
effort may depend on the explicit shape of the gamma matrices.
Therefore, in dependence on the physical problem under consideration
one may ask whether it is possible to find a coordinate system in
which the gamma matrices assume a particularly convenient shape for
some calculational purpose. The detailed requirements certainly 
may depend on the purpose. From such a problem, recently we have 
been led to ask ourselves whether it is possible to find an 
irreducible representation of the real Clifford algebra
C(3,1) which is particularly symmetric with
respect to the index $\mu$ of the gamma matrices $\gamma^\prime_\mu$.
Indeed, it is possible to find an orthogonal transformation which 
transforms the gamma matrices (\ref{A2a})-(\ref{A2d}) into the 
following expressions which are obviously particularly symmetric
with respect to the index of the gamma matrices $k = 1, 2, 3$
(${\bf 1}$ and ${\bf 0}$ are the $2 \times 2$ unit and null
matrices, respectively; $\varphi_0$ is some arbitrary real 
constant; cf.\ sect.\ 5). 
\begin{eqnarray}
\label{A4a}
\gamma^\prime_k&=&
\frac{1}{\sqrt{3}}
\left(
\begin{array}{*{2}{c}}
{\bf 1}&{\bf F}_k\\
{\bf F}_k&{\bf -1}
\end{array}
\right)\ \ \ ,\hspace{1cm}
{\bf F}_k\ =\ 
\left(
\begin{array}{*{2}{c}}
f(-\varphi_k)&f(\varphi_k)\\
f(\varphi_k)&-f(-\varphi_k)
\end{array}
\right)\\[0.3cm]
&&\hspace{0.5cm}f(\varphi)\ =\ \cos\varphi\ +\ \sin\varphi\ =\
\sqrt{2}\cos\left(\varphi-\frac{\pi}{4}\right)\\[0.3cm]
&&\label{A4ac}
\hspace{0.5cm}\varphi_k\ = \ \varphi(k)\ =\ 
\varphi_0\ + \ \frac{2\pi}{3}\ k\\[0.3cm]
\label{A4b}
\gamma^\prime_4&=&
\left(
\begin{array}{*{2}{c}}
{\bf 0}&{\bf -1}\\
{\bf 1}&{\bf 0}
\end{array}
\right)
\end{eqnarray}
As a byproduct, from the above expressions one obtains the following
compact formula for transformed Pauli matrices (irreducible matrix
representations of the complex Clifford algebra C(3,0); cf.\ 
Appendix B).
\begin{eqnarray}
\label{A5a}
\sigma^\prime_k&=&
\frac{1}{\sqrt{3}}
\left(
\begin{array}{*{2}{c}}
1&\sqrt{2}\ {\rm e}^{-i\varphi_k}\\
\sqrt{2}\ {\rm e}^{i\varphi_k}&-1
\end{array}
\right)
\end{eqnarray}
It is the purpose of the present article to systematically 
derive the above expressions relying on certain information 
not applied previously within the present context.
The discussion is accompanied by references to the 
relevant but scattered literature.\\

Our considerations will be guided by the following idea. 
Related to the Clifford algebra C(3,0), it should be possible
to find an expression for the set of the gamma matrices 
$\gamma^\prime_k$, $k = 1, 2, 3$
which is particularly symmetric with respect to the index $k$.
We approach the problem by noting that each gamma matrix $\gamma_k$ has 
2 two-dimensional eigenspaces related to the eigenvalues
$\rho = 1$ and $\rho = -1$ (which are orthogonal to 
each other). Any coordinate 
system in $V$ stands in a certain geometric relation to all
the eigenspaces of the gamma matrices whose mutual relation
is an invariant under any transformation (\ref{A3}). 
Now, the idea consists
in finding such a coordinate system in $V$ with respect to
which all the eigenspaces of the gamma matrices lie in a 
particularly symmetric way. Then, one may expect that the 
explicit expressions for the gamma matrices 
$\gamma^\prime_k$ reflect this 
symmetry. Therefore, in the next section we start with some
observations concerning the eigenspaces of the generators of 
the Clifford algebra C(3,0) (more precisely, in using this 
term we always mean the generators of its irreducible representations).\\

\section{Isoclinic 2-planes in ${\bf R_4}$}

To begin with, let us discuss some aspects of the geometry
of 2-planes in the affine space ${\bf R}_4$ which we also
denote by $V$ for simplicity. We restrict
our consideration to 2-planes containing the point 
${\bf x} = (0,0,0,0)$ (i.e.\ to the Grassmann manifold G(2,4),
for a related review see \cite{bori}).
We will rely here on the general multidimensional 
matrix formalism presented in \cite{roze},
ch.\ 3, \S 3 (also see \cite{roze2}, ch.\ III, \S 3.3)
which we specialize to ${\bf R}_4$. 
In the following we will start
with some material which provides the necessary information
on those aspect of the formalism of \cite{roze}, \cite{roze2} which is 
relevant for the present paper.\\

For our purposes, a point ${\bf x}$ of a given 2-plane $A$ can be described
in terms of the equation
\begin{eqnarray}
\label{B1}
{\bf x}&=&{\bf A\ t}
\end{eqnarray}
where ${\bf A}$ is a $4\times 2$ matrix whose two columns are given by the 
coordinates of two linearly independent vectors spanning the 2-plane $A$ while
${\bf t}$ is the two-component vector of the coordinates of the point
${\bf x}\in A$. Two 2-planes $A$ and $B$ can intersect in $V$
in various ways. In order to study their relation,
to each pair of vectors ${\bf x}\in A$, 
${\bf y} \in B$ the angle they enclose can be calculated.
Once a vector ${\bf x}\in A$ is fixed, for any arbitrary vector 
${\bf y} \in B$ the angle enclosed assumes values between some
$\alpha_0 \ge 0$ and $\pi/2$. In general, $\alpha_0$ may lie 
between some minimal and some maximal value -- 
the so-called {\it stationary angles} $\alpha_{min}$,
$\alpha_{max}$ -- which are characteristic for the 
geometry of the pair of 2-planes $A$, $B$.
Now, from an extremum principle a $2 \times 2$ matrix 
\begin{eqnarray}
\label{B2}
{\bf W}&=& 
\left( {\bf A}^T {\bf A}\right)^{-1}
\left( {\bf A}^T {\bf B}\right)
\left( {\bf B}^T {\bf B}\right)^{-1}
\left( {\bf B}^T {\bf A}\right)
\end{eqnarray}
can be constructed\footnote{Eq.\ (3.109) in \cite{roze}, 
sect.\ 3.3.15, p.\ 107 and eq.\ (3.13) in \cite{roze2}, 
sect.\ 3.3.3, p.\ 179;
this matrix has also been considered in \cite{afri}, sect.\ 5,
p.\ 807, \cite{zass}, \cite{beni}, sect.\ 4, p.\ 1195, and 
\cite{lant}, p.\ 242, eq.\ (3).
An equivalent definition based on a parameterization like
(\ref{C1a}), (\ref{C1b}) is given in \cite{wong1}, pt.\ I,
sect.\ 2, p.\ 13 and in \cite{wong2}, sect.\ 1-3,
p.\ 16, theorem 1-3.5. Further useful information with 
respect to the angles between planes can be found in \cite{shir},
\S 31, 5., p.\ 392, \cite{wong3}, \cite{ivan}, ch.\ 1, p.\ 23,
\cite{watk}, sect.\ 7.5, p.\ 421,
and \cite{ipse}
(also in \cite{woo} which, however, overlaps 
with \cite{wong2} and \cite{shir}). Also note
\cite{davi} and references therein (sect.\ 3, p.\ 18, bibliographical
note). Finally, for the convenience of the reader
we would like to mention some further 
recent references dealing with the angles between subspaces 
\cite{wedi}-\cite{jian}.}
for whose eigenvalues $w_1$ and $w_2$ the equations
\begin{eqnarray}
\label{B3a}
w_1&=&\cos^2\alpha_{max}\ \ \ ,\\[0.3cm]
\label{B3b}
w_2&=&\cos^2\alpha_{min}
\end{eqnarray}
apply. If the 2-planes $A$, $B$ are given by means of eq.\ (\ref{B1})
in terms of two orthonormal vectors each, eq.\ (\ref{B2})
simplifies to the form\footnote{Eq.\ (3.110) in \cite{roze}, 
sect.\ 3.3.15, p.\ 108 and eq.\ (3.15) in \cite{roze2}, 
sect.\ 3.3.3, p.\ 179; this matrix has also been considered in
\cite{flan}, p.\ 136, \cite{vent}, sect.\ 2,
\cite{reich}, ch.\ V, \S 3., p.\ 283,
\cite{vitn}, p.\ 416, eq.\ ($5^\prime$),
and \cite{dege}, sect.\ 1; also see \cite{watk}, sect.\ 7.5,
corollary 7.5.9, p.\ 425.}
\begin{eqnarray}
\label{B4}
{\bf W}&=& 
\left( {\bf A}^T {\bf B}\right)\left( {\bf B}^T {\bf A}\right)\ \ \ .
\end{eqnarray}
If the matrix ${\bf W}$ is proportional to the unit matrix
(i.e.\ $w_1 = w_2 = w$)
\begin{eqnarray}
\label{B5}
{\bf W}&=& 
w\ {\bf 1}\ \ \ ,
\end{eqnarray}
the 2-planes $A$ and $B$ are said to be
(mutually) {\it isoclinic}\footnote{\cite{roze}, 
sect.\ 3.3.16, p.\ 109. Eq.\ (\ref{B5}) has also been
considered in \cite{lemm}, p.\ 99, 2.3; a related equation
can be found in \cite{brun2}, sect.\ 4, p.\ 144
and in \cite{brun1}, sect.\ 3, p.\ 533, eq.\ (3.2);
also see \cite{guns}, sect.\ 2, p.\ 299, eq.\ (2.3), and
\cite{shap}, sect.\ 1, p.\ 481, (1.2).
Instead of the term isoclinic also the terms {\it isocline} \cite{mann}
and {\it isogonal} \cite{afri} have been used.}. 
Then, to each vector ${\bf x}\in A$ a unique 
line in $B$ exists (determined by the orthogonal projection of 
${\bf x}$ onto $B$) which encloses with ${\bf x}$
the (stationary) angle $\alpha = \arccos\sqrt{w}$ \footnote{For 
some further information 
with respect to isoclinic 2-planes the reader is referred to 
\cite{wong2} (which, however, does not contain any references)
and \cite{mann}, in particular ch.\ III, sect.\ III and ch.\ IV,
sect.\ VI. Manning \cite{mann} cites as his main source a 
comprehensive article by Stringham \cite{stri} which contains a 
list of related references from the nineteenth 
century literature. A comprehensive overview of the 
literature on multidimensional Euclidean geometry 
before 1911 can be found in \cite{somm3}.
Finally, the subject of isoclinic 2-planes has also been dealt with
in the monographs \cite{fors}, \cite{scho} and \cite{somm2}
(the sequence indicates the depth of the discussion with
\cite{fors} being the most comprehensive one among these three titles).}.
Finally, we would like to mention that under some natural 
bijection between ${\bf R}_4$ and ${\bf C}_2$ 
($(z_1,z_2) = (x_1 + i x_2, x_3 + i x_4) \in {\bf C}_2$, 
$(x_1,x_2,x_3,x_4) \in {\bf R}_4$) two isoclinic 
2-planes in ${\bf R}_4$ correspond to two lines 
through the origin in 
${\bf C}_2$ (\cite{wong2}, sect.\ 1-7, p.\ 51, theorem 1-7.4).\\

Now, the above formalism can be used to analyse the geometry
of the set of 6 two-dimensional eigenspaces of the generators of 
the Clifford algebra C(3,0) (i.e. more precisely, the generators
of its irreducible representation). After some calculation using 
e.g.\ the explicit representations of the gamma matrices
(\ref{A2a})-(\ref{A2c})
one finds that all their six eigenspaces are pairwise isoclinic
2-planes (some choice for the matrices ${\bf A}$ describing the 
eigenspaces is given in Appendix A). 
Of course, the two eigenspaces of a given gamma matrix 
$\gamma_k$ are orthogonal to each other. But, any other two
eigenspaces are pairwise isoclinic with an (stationary) angle 
$\alpha = \pi/4$. Consequently, we can find, at maximum, 
a set of three eigenspaces of the gamma matrices 
$\gamma_k$, $k = 1, 2, 3$, whose elements are 
pairwise isoclinic with the angle $\pi/4$ \footnote{This is the maximal
possible number of equi-isoclinic 2-planes with angle $\pi/4$
in ${\bf R}_4$ in general, cf.\ \cite{lemm}, sect.\ 5,
(5.1); incidentally, note some generalizations of this result
obtained in \cite{hogg1}-\cite{hogg3} for 
complex and quaternionic spaces $V$.}. 
Such a set of 2-planes is called 
an {\it equiangular frame} (\cite{wong1}, pt.\ I, sect.\ 5, p.\ 40). 
With respect to the aim of the present paper, 
in the following we will just be interested in such sets.\\

\section{The Clifford algebra C(3,0) and equiangular\hfill\break
frames}

We begin this section with some necessary information taken 
from \cite{wong1}\footnote{The results of this paper have been
reformulated from a different point of view in \cite{tyrr}.}
and specialized to the present needs
(in the following the term `adapted quote' always means that
the original text is quoted exactly except that any reference 
to the general multidimensional space ${\bf R}_{2n}$
has been specialized to ${\bf R}_4$).
The following definition will be used: 
``A {\it set of mutually isoclinic 2-planes} in ${\bf R}_4$
is characterized by the property that every two 2-planes of the 
set are isoclinic with each other. A set of mutually isoclinic
2-planes in ${\bf R}_4$ is called a {\it maximal} set if it 
is not subset of a larger set of mutually isoclinic 2-planes''
(This is an adapted quote from \cite{wong1}, pt.\ I, sect.\ 3, p.\ 
19)\footnote{In \cite{wong4} the shorter term {\it isoclinic set of 
2-planes} is used instead of the term `maximal set of 
mutually isoclinic 2-planes'. As we mainly rely on \cite{wong1},
in the present article we use the original term.
Furthermore, every maximal set of mutually isoclinic 2-planes is also
a {\it maximal normal set of 2-planes} \cite{wong4}, \cite{yiu}. 
The term {\it normal set} denotes a set of 2-planes which are 
pairwise either orthogonal or normally related 
to each other. Two 2-planes $A$, $B$ are {\it normally related} if
$A \bigcap B = A \bigcap B^{\perp} = \{0\}$. Finally, 
the relation of maximal sets of mutually isoclinic 2-planes to the Hopf map
has been considered in \cite{wong5}.}.
In order to make contact with the formalism used in \cite{wong1}
which we will rely on in the further discussion
we need to rewrite the defining equation (\ref{B1}) for a 2-plane $A$
in one of the following two (alternative) ways. 
\begin{eqnarray}
\label{C1a}
{\bf x}_{(3,4)}&=&\tilde{\bf A}\ {\bf x}_{(1,2)}\ \ ,\ \ \
\tilde{\bf A}\ =\ \bar{\bar{\bf A}}
\left(\bar{\bf A}\right)^{-1}\\[0.3cm]
\label{C1b}
{\bf x}_{(1,2)}&=&\tilde{\tilde{\bf A}}\ {\bf x}_{(3,4)}\ \ ,\ \ \
\tilde{\tilde{\bf A}}\ =\ \bar{\bf A}
\left(\bar{\bar{\bf A}}\right)^{-1}
\end{eqnarray}
Here, the notation 
${\bf x}_{(1,2)} = (x_1, x_2)^T$, ${\bf x}_{(3,4)} = (x_3, x_4)^T$ is used
and the $2\times 2$ matrices $\bar{\bf A}$, $\bar{\bar{\bf A}}$
are related to the matrix ${\bf A}$ the following way.
\begin{eqnarray}
\label{C2}
{\bf A}&=&
\left(
\begin{array}{*{1}{c}}
\bar{\bf A}\\ 
\bar{\bar{\bf A}}
\end{array}
\right)
\end{eqnarray}
Eq.\ (\ref{C1a}) [(\ref{C1b})] is valid for any 2-plane which 
is isoclinic but not identical to the 2-plane $O_{(3,4)}$:
${\bf x}_{(1,2)} = 0$ [$O_{(1,2)}$: ${\bf x}_{(3,4)} = 0$]
(this entails that the 2-plane $A$ intersects 
the 2-plane $O_{(3,4)}$ [$O_{(1,2)}$] in the 
point  ${\bf x} = (0,0,0,0)$  only and, therefore, ensures the
invertibility of $\bar{\bf A}$ [$\bar{\bar{\bf A}}$]).\\

According to Wong (\cite{wong1},
pt.\ I, sect.\ 7, p.\ 54, theorem 7.2; also see \cite{wong2},
sect.\ 1-7, p.\ 43), every maximal set of 
mutually isoclinic 2-planes in ${\bf R}_4$ is of dimension 2 
and is congruent (i.e.\ related by an orthogonal
transformation in ${\bf R}_4$) to the maximal set given by the 
equation
\begin{eqnarray}
\label{C3a}
{\bf x}_{(3,4)}&=&\tilde{\bf B}(\lambda_0,\lambda_1) \ {\bf x}_{(1,2)}\ =\ 
\left[ \lambda_0\tilde{\bf B}_0\ +\ 
\lambda_1\tilde{\bf B}_1\right]\ {\bf x}_{(1,2)}\ \ ,\\[0.3cm]
\label{C3b}
\tilde{\bf B}_0&=&\left(
\begin{array}{*{2}{c}}
-1&0\\ 
0&1
\end{array}\right) \ \ ,\ \ \ 
\tilde{\bf B}_1\ =\ \left(
\begin{array}{*{2}{c}}
0&1\\ 
1&0
\end{array}
\right)
\end{eqnarray}
or,
\begin{eqnarray}
\label{C3c}
{\bf x}_{(1,2)}&=&\tilde{\tilde{\bf B}}(\lambda_0,\lambda_1) 
\ {\bf x}_{(3,4)}\ \ ,\\[0.3cm]
\label{C3d}
\tilde{\tilde{\bf B}}(\lambda_0,\lambda_1)&=&
\tilde{\bf B}(\lambda_0,\lambda_1)^{-1}\ =\ 
\frac{1}{\lambda^2_0 + \lambda^2_1}\ 
\tilde{\bf B}(\lambda_0,\lambda_1)\ =\ 
\tilde{\bf B}(\lambda^\prime_0,\lambda^\prime_1)\ \ ,\\[0.3cm]
&&\ \ \ \ \ \lambda^{\prime}_n\ =\ 
\frac{\lambda_n}{\lambda^2_0 + \lambda^2_1}\ \ ,\ \ \  
n\ =\ 1, 2\ \ ,\nonumber
\end{eqnarray}
where $\lambda_0$, $\lambda_1$ are two real 
parameters\footnote{The corresponding 
expression of Wong \cite{wong1} is related to eq.\ (\ref{C3a}) by an
orthogonal transformation (inversion) 
$(x_1,x_2,x_3,x_4)\rightarrow (- x_1,x_2,x_3,x_4)$.
Eq.\ (\ref{C3a}) has been re-derived in \cite{reye}
by means of fairly elementary considerations.
$\tilde{\bf B}(\lambda_0,\lambda_1)$ can be written as
$\tan\theta\ \tilde{\bf D}$ ($\tan\theta = \sqrt{\lambda^2_0 + \lambda^2_1}$), 
where $\theta$ is the angle between the 
2-planes $O_{(1,2)}$, $B$, and $\tilde{\bf D}$ is an orthogonal matrix
($\tilde{\bf D} \in O(2)$) \cite{brun1}.}.
Both of the 2-planes $O_{(1,2)}$: ${\bf x}_{(3,4)} = 0$ and 
$O_{(3,4)}$: ${\bf x}_{(1,2)} = 0$ belong to this maximal set
(\cite{wong1}, pt.\ I, sect.\ 2, p.\ 16, lemma 2.2).
Eqs.\ (\ref{C3a}), (\ref{C3c}) entail that the matrix ${\bf B}$
to be inserted in the corresponding eq.\ (\ref{B1}) reads, e.g.,
(we have chosen particularly simple expressions)
\begin{eqnarray}
\label{C4a}
{\bf B}(\lambda_0,\lambda_1)&=&
\frac{1}{\sqrt{1 +\lambda^2_0 +\lambda^2_1}}
\left(
\begin{array}{*{1}{c}}
{\bf 1}\\ 
\tilde{\bf B}(\lambda_0,\lambda_1)
\end{array}
\right)\ \ \ ,
\end{eqnarray}
or
\begin{eqnarray}
\label{C4b}
{\bf B}(\lambda^\prime_0,\lambda^\prime_1)&=&
\frac{1}{\sqrt{1 +\lambda^{\prime\; 2}_0 +\lambda^{\prime\; 2}_1}}
\left(
\begin{array}{*{1}{c}}
\tilde{\bf B}(\lambda^\prime_0,\lambda^\prime_1)\\
{\bf 1}
\end{array}
\right)\ \ \ .
\end{eqnarray}
Furthermore, Wong finds that (adapted quote)
``in ${\bf R}_4$, any maximal set of mutually isoclinic 2-planes
which contains the 2-plane $O_{(1,2)}$ corresponds to a linear
subspace of the linear space of all $2\times 2$ matrices''
(\cite{wong1}, pt.\ I, sect.\ 3, p.\ 20, lemma 3.2).
Now, in this 2-dimensional subspace a matrix basis can be chosen 
such a way that the 2-planes described by the 
elements of the basis and the 2-plane $O_{(1,2)}$ (or $O_{(3,4)}$)
form an equiangular frame (\cite{wong1}, pt.\ I, sect.\ 3, p.\ 24, 
lemma 3.3 and p.\ 40). As one may convince oneself easily by
means of the explicit expressions given in Appendix A, each equiangular
frame built from the eigenspaces of the gamma matrices contains a 
basis of one and the same maximal set of mutually isoclinic 2-planes.\\

For the purpose of the present paper 
it appears to be useful to consider two disjoint equiangular
frames $\Omega$ connected with the gamma matrices (\ref{A2a})-(\ref{A2c}) -- 
one ($\Omega_1$) related to the three eigenspaces to the eigenvalue 
$\rho = 1$, and the other one ($\Omega_{-1}$) related to the three eigenspaces
to the eigenvalue $\rho = -1$. The following theorem by Wong 
will be helpful then ($\Phi$ is any maximal set of mutually isoclinic 
2-planes in ${\bf R}_4$; the following is an 
adapted quote; the indices have also been changed to conform
to the notation used in the present article): ``If the angles 
between any 2-plane of $\Phi$ and the three 2-planes of an 
equiangular frame are $\theta_k$ $(1 \le k \le 3)$, then
\begin{eqnarray}
\label{C5}
\cos^2 2\theta_1\ +\ \cos^2 2\theta_2\ +\ \cos^2 2\theta_3&=&1\ \ .
\end{eqnarray}
Conversely, given any set of three angles $\theta_k$ $(1 \le k \le 3)$
such that $0 \le \theta_k \le \pi$ and $\sum \cos^2 2 \theta_k = 1$,
then there exists a unique 2-plane isoclinic to each of the three
2-planes of a given equiangular frame, making angles $\theta_k$ with
them, and this 2-plane belongs to $\Phi$'' (\cite{wong1}, pt.\ I, 
sect.\ 5, p.\ 41, theorem 5.3 (b)). From this insight we
conclude that, obviously, to each 
equiangular frame $\Omega_1$ [$\Omega_{-1}$] two uniquely
determined 2-planes $A_{1\pm}$ 
[$A_{-1\pm}$] exist which lie in a particularly symmetric way 
(isoclinic) relative to the elements of $\Omega_1$ [$\Omega_{-1}$]. 
For $A_{1\pm}$, $A_{-1\pm}$ it holds 
\begin{eqnarray}
\label{C6}
\theta_1&=&\theta_2\ =\ \theta_3\ =\ \theta_{sym}\ \ ,\ \ \ 
\cos 2\theta_{sym}\ =\ \pm \frac{1}{\sqrt{3}}\ \ .
\end{eqnarray}
For the corresponding eigenvalue of the matrix ${\bf W}$, eq.\ (\ref{B5}),
one obtains
\begin{eqnarray}
\label{C7}
w&=&\cos^2 \theta_{sym}\ =\ \frac{1}{2}\ \left(1\ +\
\cos 2\theta_{sym}\right)\ =\ 
\frac{1}{2}\ \left(1\ \pm \frac{1}{\sqrt{3}}\right)\ = \ w_\pm\ \ .
\end{eqnarray}
The two different values of $\theta_{sym}$ (and $w$) will not cause
any major difference in the following considerations as both cases
are related by a simple permutation of the indices of the gamma 
matrices.\\

\section{Change of the coordinate system}

We may now set out to determine the position of the 2-planes
$A_{1\pm}$, $A_{-1\pm}$ using the formulae given in the two preceding
sections. For the 2-planes $A_{1\pm}$, $A_{-1\pm}$ we can apply a 
general Ansatz according to eqs.\ (\ref{C3a}), (\ref{C3c}), 
(\ref{C4a}), (\ref{C4b}) and 
calculate the eigenvalue of the matrix ${\bf W}$ for each of the 
three pairs given by one of the elements of the equiangular 
frame $\Omega_1$ [$\Omega_{-1}$] and $A_{1\pm}$ [$A_{-1\pm}$]. 
For each eigenvalue $\rho$ of the gamma matrices 
(\ref{A2a})-(\ref{A2c}), this leads to a set of three 
equations for the parameters $\lambda_0$, $\lambda_1$ which have to be
solved simultaneously taking into account eq.\ (\ref{C7}).
These equations read for $\rho = 1$ (in sequence for the indices $k = 1$, 
$k = 2$ and $k =3$ of the gamma matrices, respectively)
\begin{eqnarray}
\label{D1a}
w_{\pm}&=&\frac{\lambda^{\prime\; 2}_0 + 
\left(1 + \lambda^\prime_1\right)^2}{
2\left(1 + \lambda^{\prime\; 2}_0 + 
\lambda^{\prime\; 2}_1\right)}\ \ ,\\[0.3cm]
\label{D1b}
w_{\pm}&=&\frac{\left(1 - \lambda^\prime_0\right)^2 + 
\lambda^{\prime\; 2}_1}{
2\left(1 + \lambda^{\prime\; 2}_0 + 
\lambda^{\prime\; 2}_1\right)}\ \ ,\\[0.3cm]
\label{D1c}
w_{\pm}&=&\frac{\lambda^{\prime\; 2}_0 + \lambda^{\prime\; 2}_1}{
1 + \lambda^{\prime\; 2}_0 + \lambda^{\prime\; 2}_1}\ \ ,
\end{eqnarray}
and for $\rho = -1$,
\begin{eqnarray}
\label{D2a}
w_\pm&=&\frac{\lambda^2_0 + \left(1 - \lambda_1\right)^2}{
2\left(1 + \lambda^2_0 + \lambda^2_1\right)}\ \ ,\\[0.3cm]
\label{D2b}
w_\pm&=&\frac{\left(1 + \lambda_0\right)^2 + \lambda^2_1}{
2\left(1 + \lambda^2_0 + \lambda^2_1\right)}\ \ ,\\[0.3cm]
\label{D2c}
w_\pm&=&\frac{\lambda^2_0\ +\ \lambda^2_1}{
1 + \lambda^2_0 + \lambda^2_1}
\end{eqnarray}
(The eqs.\ (\ref{D1a})-(\ref{D1c}) [(\ref{D2a})-(\ref{D2c})] have 
been derived using the expressions given in Appendix A 
and eq.\ (\ref{C4b}) [(\ref{C4a})].).
The solution of the above equations reads for $\rho = 1$
\begin{eqnarray}
\label{D3}
\lambda^\prime_0&=&-\lambda^\prime_1\ =\ - \lambda_\pm\ \ ,
\end{eqnarray}
and for $\rho = -1$
\begin{eqnarray}
\label{D4}
\lambda_0&=&-\lambda_1\ =\ \lambda_\pm\ \ \ .
\end{eqnarray}
Here, 
\begin{eqnarray}
\label{D4b}
\lambda_\pm &=& \pm \sqrt{3}\ w_\pm
\end{eqnarray}
which entails
\begin{eqnarray}
\label{D4c}
2 \lambda_\pm \lambda_\mp &=& -1\ \ .
\end{eqnarray}

Now, we may assume that the explicit representations for the 
gamma matrices (\ref{A2a})-(\ref{A2d}) are related to the 
natural basis in $V$ from which two pairs of basis vectors can be
selected which define the orthogonal 2-planes $O_{(1,2)}$, $O_{(3,4)}$. 
In order to obtain a particularly symmetric representation 
for the gamma matrices it appears to be advantageous now to 
go over to an orthonormal basis from which two pairs 
of basis vectors can be chosen which define the orthogonal 
2-planes $A_{1\pm}$, $A_{-1\pm}$. 
This change of the basis in $V$ is associated with
an orthogonal transformation $O$ in $V$ which transforms 
the gamma matrices in accordance with eq.\ (\ref{A3}).
We start by choosing an appropriate orthonormal basis in $V$
from which the matrices ${\bf A}_{1\pm}$, ${\bf A}_{-1\pm}$ 
describing the 2-planes $A_{1\pm}$, $A_{-1\pm}$ can be built (we simply
insert the solutions (\ref{D3}), (\ref{D4})
into the eqs.\ (\ref{C4b}), (\ref{C4a}), respectively).
\begin{eqnarray}
\label{D5a}
{\bf A}_{1\pm}&=&
\frac{1}{\sqrt{1 + 2\lambda^2_{\pm}}}
\left(
\begin{array}{*{2}{c}}
\lambda_\pm&\lambda_\pm\\ 
\lambda_\pm&-\lambda_\pm\\
1&0\\
0&1
\end{array}
\right)\\[0.3cm]
\label{D5b}
{\bf A}_{-1\pm}&=&
\frac{1}{\sqrt{1 + 2 \lambda^2_{\pm}}}
\left(
\begin{array}{*{2}{c}}
1&0\\
0&1\\
-\lambda_\pm&-\lambda_\pm\\
-\lambda_\pm&\lambda_\pm
\end{array}
\right)
\end{eqnarray}
One immediately recognizes that the 
2-planes $A_{1\pm}$, $A_{-1\pm}$ are orthogonal to each other.
Furthermore, by virtue of eq.\ (\ref{D4c}) it 
holds $A_{1\pm} = A_{-1\mp}$.
Of course, the above choice for the matrices
${\bf A}_{1\pm}$, ${\bf A}_{-1\pm}$ is not unique and any orthonormal 
basis which is related to the basis used in the above equations by a rotation
within the 2-planes $A_{1\pm}$, $A_{-1\pm}$ is equally well suited. 
In fact, further below we will use exactly this freedom to 
obtain our final result (\ref{A4a})-(\ref{A4b}).\\

The transition from the natural basis in $V$ which is related
to the 2-planes $O_{(1,2)}$, $O_{(3,4)}$ to the basis 
which is given in terms of eqs.\ (\ref{D5a}), (\ref{D5b}) and which is
related to the 2-planes $A_{1\pm}$, $A_{-1\pm}$ is described by
the orthogonal transformation $O_\pm$
\begin{eqnarray}
\label{D6}
O_\pm&=&
\frac{1}{\sqrt{1 + 2\lambda^2_{\pm}}}
\left(
\begin{array}{*{4}{c}}
\lambda_\pm&\lambda_\pm&1&0\\ 
\lambda_\pm&-\lambda_\pm&0&1\\
1&0&-\lambda_\pm&-\lambda_\pm\\
0&1&-\lambda_\pm&\lambda_\pm\\
\end{array}
\right)
\end{eqnarray}
which leads via $\gamma^{\prime\prime}_\mu = O_\pm \gamma_\mu O^T_\pm$ 
to the correspondingly 
transformed expressions for the gamma matrices $\gamma_\mu$
(of course, for our choice (\ref{D6}) it holds $O_\pm = O^T_\pm $).
After some algebra (taking into account eq.\ (\ref{D4c}))
one finds
\begin{eqnarray}
\label{D7a}
\gamma^{\prime\prime}_{1\pm}&=&-\ \gamma^{\prime\prime}_{2\mp}\ =\ 
\pm\frac{1}{\sqrt{3}}
\left(
\begin{array}{*{4}{c}}
1&0&-\lambda_\pm&-\lambda_\mp\\ 
0&1&-\lambda_\mp&\lambda_\pm\\
-\lambda_\pm&-\lambda_\mp&-1&0\\
-\lambda_\mp&\lambda_\pm&0&-1
\end{array}
\right)\ \ ,\\[0.3cm]
\label{D7c}
\gamma^{\prime\prime}_{3\pm}&=&
\pm\frac{1}{\sqrt{3}}
\left(
\begin{array}{*{4}{c}}
1&0&1&1\\ 
0&1&1&-1\\
1&1&-1&0\\
1&-1&0&-1
\end{array}
\right)\ \ ,\\[0.3cm]
\label{D7d}
\gamma^{\prime\prime}_4&=&
\left(
\begin{array}{*{4}{c}}
0&0&-1&0\\ 
0&0&0&-1\\
1&0&0&0\\
0&1&0&0
\end{array}
\right)\ \ .
\end{eqnarray}
 From eq.\ (\ref{D7a}) one immediately recognizes that the two 
cases differing by the sign in eq.\ (\ref{C6}) are related 
to each other by a permutation of the gamma matrices with 
the indices $k = 1$ and $k = 2$.\\

\section{Residual rotations}

Although in the preceding section we have performed the 
transformation to a coordinate system which lies in a particularly
symmetric way with respect to the equiangular frames 
$\Omega_1$, $\Omega_{-1}$ built from the eigenspaces of the 
gamma matrices, at first glance the transformed expressions
(\ref{D7a}), (\ref{D7c}) do not seem to exhibit any particular 
symmetry with respect to the index $k = 1, 2, 3$ of the gamma matrices.
However, the expected symmetry is there and we are going to reveal
it now. Let us remind ourselves that the choice of the new 
basis (coordinate system) was not unique and we have disregarded
for the moment the remaining freedom to perform rotations within 
the 2-planes $A_{1\pm}$, $A_{-1\pm}$. Any such rotation can be 
described by the orthogonal transformation
\begin{eqnarray}
\label{E1}
O(\beta_1,\beta_{-1})&=&
\left(
\begin{array}{*{4}{c}}
\cos\beta_1&-\sin\beta_1&0&0\\ 
\sin\beta_1&\cos\beta_1&0&0\\
0&0&\cos\beta_{-1}&-\sin\beta_{-1}\\ 
0&0&\sin\beta_{-1}&\cos\beta_{-1}
\end{array}
\right)
\end{eqnarray}
where $\beta_1$ and $\beta_{-1}$ are the independent 
rotation angles within the orthogonal 2-planes $A_{1\pm}$ 
and $A_{-1\pm}$, respectively (for the sake of completeness
we mention that in addition to the above rotations an
inversion within one of the 2-planes $A_{1\pm}$, $A_{-1\pm}$
may be considered). Again, we can write
down the further transformed gamma matrices
$\gamma^\prime_\mu = O(\beta_1,\beta_{-1})\gamma^{\prime\prime}_\mu
O(\beta_1,\beta_{-1})^T$. For brevity,
we give the relatively simple expressions for $\gamma^\prime_{3\pm}$ and 
$\gamma^\prime_4$ only.
\begin{eqnarray}
\label{E2a}
\gamma^\prime_{3\pm}(\varphi)&=&
\pm\frac{1}{\sqrt{3}}
\left(
\begin{array}{*{4}{c}}
1&0&f(-\varphi)&f(\varphi)\\ 
0&1&f(\varphi)&-f(-\varphi)\\
f(-\varphi)&f(\varphi)&-1&0\\
f(\varphi)&-f(-\varphi)&0&-1
\end{array}
\right)\\[0.3cm]
\label{E2b}
\gamma^\prime_4(\bar{\varphi})&=&
\left(
\begin{array}{*{4}{c}}
0&0&-\cos\bar{\varphi}&\sin\bar{\varphi}\\ 
0&0&-\sin\bar{\varphi}&-\cos\bar{\varphi}\\
\cos\bar{\varphi}&\sin\bar{\varphi}&0&0\\
-\sin\bar{\varphi}&\cos\bar{\varphi}&0&0
\end{array}
\right)
\end{eqnarray}
Here, $\varphi = \beta_1 +\beta_{-1}$ and 
$\bar{\varphi} = \beta_1 -\beta_{-1}$. The gamma matrices
$\gamma^\prime_{k\pm}$, $k = 1, 2, 3$, do not depend on $\bar{\varphi}$
while $\gamma^\prime_4$ does not depend on $\varphi$.
The function $f$ is given by
\begin{eqnarray}
\label{E3}
f(\varphi)&=&\cos\varphi\ +\ \sin\varphi
\ =\ \sqrt{2}\cos\left(\varphi -\frac{\pi}{4}\right)\ \ .
\end{eqnarray}
Symmetry considerations now suggest
that any set of (three) rotations
$O(\beta_1,\beta_{-1})$ among whose elements 
$\varphi = \beta_1 +\beta_{-1}$ 
changes by a multiple of $2\pi/3$ (mod $2\pi$) will lead to a set of three
gamma matrices with the indices $k = 1, 2, 3$. 
Consequently, in order to describe this set we can write
\begin{eqnarray}
\label{E4}
\varphi(k)&=&\varphi_0\ +\ \frac{2\pi}{3}\ k\ =\ \varphi_k
\end{eqnarray}
where $\varphi_0$ is some real constant. 
Any three gamma matrices given by eqs.\ (\ref{E2a}), (\ref{E3}) and (\ref{E4})
can be chosen to serve as an irreducible representation of the real 
Clifford algebra C(3,0). If we choose $\varphi_0 = 0$, eqs.\ 
(\ref{E2a}), (\ref{E3}) and (\ref{E4}) exactly reproduce the set of
gamma matrices (\ref{D7a}), (\ref{D7c}), i.e.
\begin{eqnarray}
\label{E5}
\gamma^\prime_{3\pm}\left(\frac{2\pi}{3}\right)&=&
\gamma^{\prime\prime}_{1\pm}\ \ ,\ \ \
\gamma^\prime_{3\pm}\left(\frac{4\pi}{3}\right)\ =\ 
\gamma^{\prime\prime}_{2\pm}\ \ .
\end{eqnarray}
Furthermore, for the sake of simplicity it seems to be convenient to 
set $\bar{\varphi} = 0$ and to vary $\varphi$ exclusively
(Such an orthogonal transformation is called a {\it Clifford 
translation} \cite{wong2}, sect.\ 2-6, p.\ 102
and has special properties. In this context, also note \cite{mebi}.).
This way the final result (eqs.\ (\ref{A4a})-(\ref{A4b}),
also see Appendix B for some related consideration) 
quoted in the introduction is obtained
(where we have omitted, for simplicity, the $\pm$ sign 
on the r.h.s.\ of eq.\ (\ref{E2a}) which relates to the 
two inequivalent irreducible representations of C(3,0)
\cite{okub2}, p.\ 1657).
The generators of the real Clifford algebra C(3,0) are found
from one of them by means of a discrete 
${\bf Z}_6\sim {\bf Z}_2\times{\bf Z}_3$ subgroup
of the orthogonal group $O(4)$ (in other words, the 
${\bf Z}_6$ subgroup realizes a permutation among the gamma
matrices).
The Clifford translation in the spinor space V with
$\beta_1 = \beta_{-1} = \pi/3$ 
corresponds to a rotation by 
$2\pi/3$ around the axis (1,1,1) in the vector space
${\bf R}_{3,0}$ associated with the Clifford algebra C(3,0)
(it is an element of the group $Spin(3)$).\\

We want to extend our discussion now to the real Clifford algebra C(3,2)
which is the largest Clifford algebra admitting an irreducible
representation by means of $4\times 4$ matrices. From 
eqs.\ (\ref{E2a}), (\ref{E5}) we can calculate the product
\begin{eqnarray}
\label{E6}
\gamma^\prime_{3\pm}(\varphi_1)
\gamma^\prime_{3\pm}(\varphi_2)
\gamma^\prime_{3\pm}(\varphi_3)&=&
\left(
\begin{array}{*{4}{c}}
0&1&0&0\\ 
-1&0&0&0\\
0&0&0&-1\\
0&0&1&0
\end{array}
\right)\end{eqnarray}
which is found to be independent of the choice of $\varphi_0$.
Allowing an arbitrary value for $\bar{\varphi}$,
$\gamma^\prime_5$ can then be calculated and reads
\begin{eqnarray}
\label{E7}
\gamma^\prime_5\ =\ \gamma^\prime_5(\bar{\varphi})
&=&\gamma^\prime_{3\pm}(\varphi_1)
\gamma^\prime_{3\pm}(\varphi_2)
\gamma^\prime_{3\pm}(\varphi_3)\gamma^\prime_4(\bar{\varphi})\ =\ 
\gamma^\prime_4\left(\bar{\varphi}-\frac{\pi}{2}\right)
\ \ ,\\[0.3cm]
&=&\left(
\begin{array}{*{4}{c}}
0&0&-\sin\bar{\varphi}&-\cos\bar{\varphi}\\ 
0&0&\cos\bar{\varphi}&-\sin\bar{\varphi}\\
\sin\bar{\varphi}&-\cos\bar{\varphi}&0&0\\
\cos\bar{\varphi}&\sin\bar{\varphi}&0&0
\end{array}
\right)\ \ ,\nonumber\\[0.3cm]
\gamma^{\prime\ 2}_5&=& -{\bf 1}\ \ .\nonumber
\end{eqnarray}
Finally, the charge conjugation operator $C$ ($C^T = - C$, 
$C\gamma^\prime_\mu C^{-1} = -\gamma^{\prime T}_\mu$)
can be given by $C = \gamma^\prime_4(\bar{\varphi})$.
In difference to the C(3,0) subalgebra of the Clifford algebra 
C(3,2), which is generated by relying on eq.\ (\ref{E4})
(a variation of $\varphi$ by $2\pi$ leads to just one copy of 
the generators of C(3,0)),
the C(0,2) subalgebra can be represented by 
$\gamma^\prime_4 = \gamma^\prime_4(\bar{\varphi})$, 
$\gamma^\prime_5 = \gamma^\prime_4(\bar{\varphi}\pm \pi/2)$
(a variation of $\bar{\varphi}$ by $2\pi$ leads to two copies of 
the generators of C(0,2))\footnote{This difference
between the Clifford subalgebras C(3,0) and C(0,2) 
is caused by the fact that 
during a full $2\pi$ turn around the axis (1,1,1) 
in the vector space ${\bf R}_{3,0}$ associated with C(3,0) 
the rectangular coordinate system is mapped 3 times onto
itself while during a full $2\pi$ turn
in the vector space ${\bf R}_{0,2}$ associated with C(0,2) 
the rectangular coordinate system is mapped 4 times onto
itself (allowing reflections). Taking into account the 
different number of generators of C(3,0) and C(0,2),
3 and 2, respectively, this leads to a natural explanation for
the difference.}. In this context, note
\begin{eqnarray}
\label{E8}
\gamma^\prime_4(\bar{\varphi})&=&-\ 
\gamma^\prime_4(\bar{\varphi} + \pi)\ \ .
\end{eqnarray}
For $\varphi = 0$, the second generator of the real Clifford 
algebra C(0,2) is obtained
from the first by means of a discrete 
${\bf Z}_8\sim ({\bf Z}_2)^3$ subgroup
of the orthogonal group $O(4)$.
A rotation (\ref{E1}) in the spinor space V with
$\beta_1 = - \beta_{-1} = \pi/4$ corresponds 
to a rotation by $\pi/2$ in the vector space
${\bf R}_{0,2}$ associated with the Clifford algebra C(0,2)
(it is an element of the group $Spin(2)$).\\

\section{Discussion}

According to Pauli's fundamental theorem \cite{paul}, \cite{good} 
any set of (in general, complex) $4\times 4$ gamma 
matrices $\gamma_\mu$, which represent the Clifford
algebra C(3,1), is related to our expressions for
$\gamma^\prime_\mu$ (eqs.\ (\ref{A4a})-(\ref{A4b})) by
means of a non-singular transformation $S$ 
($\gamma_\mu = S \gamma^\prime_\mu S^{-1}$). Therefore,
any such set can, in principle, be written in a
form analogous to eqs.\ (\ref{A4a})-(\ref{A4b})
(of course, in general such a representation may 
look fairly cumbersome). It is clear, that this
consideration of the (complex) Clifford algebra C(3,1) immediately
carries over with little change to the Clifford algebra 
C(1,3) and does not require any further special 
investigation. Furthermore, it seems natural to expect
that the discussion of the real Clifford algebra
C(3,1) performed in the present paper can appropriately 
be generalized also to other 
Clifford algebras. Of course, the simpler and rather trivial 
case of the real Clifford algebra C(2,1)
which is presented in Appendix C carries 
the traces of the structures found for C(3,0). 
On the other hand, one should expect that these structures 
themselves are also traces of more general 
structures of Clifford algebras which contain C(3,0) as
a subalgebra. Let us emphasize at this point 
that the mathematical tools we have 
relied on in sects.\ 2 and 3 are not specific to the 
present case (although, we have specialized them to
the present case, for simplicity) and they can also be used 
in more general situations. As interesting as this may
be, it goes far beyond the purpose of the present
study and, therefore, will not be investigated here.\\

\vspace{1.5cm}

\noindent
{\bf Acknowledgements}\\

\noindent
I am grateful to B.\ A.\ Rosenfeld for kindly informing me of 
ref.\ \cite{roze2}.
The present work has been performed under the EC Training 
and Mobility of Researchers Program,
return fellowship contract no.\ ERBFMBICT961197.

\newpage
\setcounter{section}{1}
\section*{Appendix A}
\renewcommand{\theequation}{\mbox{\Alph{section}.\arabic{equation}}}
\setcounter{equation}{0}

In this Appendix we give some explicit expressions for the matrices
${\bf A}_{k,\rho}$ which define via eq.\ (\ref{B1}) 
the eigenspace (i.e.\ the 2-plane $A_{k,\rho}$) of the gamma
matrix $\gamma_k$, $k = 1, 2, 3$, to the eigenvalue $\rho = 1, -1$.
We rely on orthonormal basis vectors for each eigenspace.

\parbox{7cm}{
\begin{eqnarray}
\label{ZA1a}
{\bf A}_{1,1}&=&
\frac{1}{\sqrt{2}}
\left(
\begin{array}{*{2}{c}}
1&0\\ 
0&1\\
0&1\\
1&0
\end{array}
\right)
\end{eqnarray}
}
\hfill
\parbox{7cm}{
\begin{eqnarray}
\label{ZA1b}
{\bf A}_{1,-1}&=&
\frac{1}{\sqrt{2}}
\left(
\begin{array}{*{2}{c}}
1&0\\ 
0&1\\
0&-1\\
-1&0
\end{array}
\right)\ \ \ \ \ 
\end{eqnarray}
}

\parbox{7cm}{
\begin{eqnarray}
\label{ZA2a}
{\bf A}_{2,1}&=&
\frac{1}{\sqrt{2}}
\left(
\begin{array}{*{2}{c}}
1&0\\ 
0&1\\
1&0\\
0&-1
\end{array}
\right)
\end{eqnarray}
}
\hfill
\parbox{7cm}{
\begin{eqnarray}
\label{ZA2b}
{\bf A}_{2,-1}&=&
\frac{1}{\sqrt{2}}
\left(
\begin{array}{*{2}{c}}
1&0\\ 
0&1\\
-1&0\\
0&1
\end{array}
\right)\ \ \ \ \
\end{eqnarray}
}

\parbox{7cm}{
\begin{eqnarray}
\label{ZA3a}
{\bf A}_{3,1}&=&
\left(
\begin{array}{*{2}{c}}
1&0\\ 
0&1\\
0&0\\
0&0
\end{array}
\right)
\end{eqnarray}
}
\hfill
\parbox{7cm}{
\begin{eqnarray}
\label{ZA3b}
{\bf A}_{3,-1}&=&
\left(
\begin{array}{*{2}{c}}
0&0\\ 
0&0\\
1&0\\
0&1
\end{array}
\right)
\end{eqnarray}
}

\noindent
 From eqs.\ (\ref{B1}), (\ref{C1a})-(\ref{C2})  
one easily recognizes that for the 2-planes 
$A_{3,1}$, $A_{3,-1}$
holds $A_{3,1} = O_{(1,2)}$, $A_{3,-1} = O_{(3,4)}$
($O_{(1,2)}$: ${\bf x}_{(3,4)} = 0$,
$O_{(3,4)}$: ${\bf x}_{(1,2)} = 0$).\\

\setcounter{section}{2}
\section*{Appendix B}
\setcounter{equation}{0}

As Pauli matrices (irreducible matrix representations of the 
complex Clifford algebra C(3,0)) play a significant role in theoretical 
physics, in this Appendix we wish to comment on the derivation 
of a particularly symmetric 
expression for these $2\times 2$ matrices by means of the approach
discussed in the main part of the paper. The standard 
expressions for the Pauli matrices read

\begin{eqnarray}
\label{ZC1}
\sigma_1&=&
\left(
\begin{array}{*{2}{c}}
0&1\\ 
1&0
\end{array}
\right)\ \ ,\ \ \ 
\sigma_2\ =\
\left(
\begin{array}{*{2}{c}}
0&-i\\ 
i&0
\end{array}
\right)\ \ ,\ \ \ 
\sigma_3\ =\
\left(
\begin{array}{*{2}{c}}
1&0\\ 
0&-1
\end{array}
\right)\ \ .\ \ 
\end{eqnarray}
In order to make contact with the main part
of the paper it turns out to be useful to represent the complex 
numbers which are entries of the matrices (\ref{ZC1})
by means of $2\times 2$ matrices using the rule
\begin{eqnarray}
\label{ZC2}
z\ =\ a + ib &\longrightarrow &
\left(
\begin{array}{*{2}{c}}
a&-b\\ 
b&a
\end{array}
\right)\ \ \ \ .
\end{eqnarray}
This leads to a set of three real $4\times 4$ matrices which are
congruent to the gamma matrices given by eq.\ (\ref{A4a}).
In order to obtain the desired final result we have to subject 
the latter gamma matrices to a further orthogonal transformation --
an inversion (mentioned below eq.\ (\ref{E1})). Then the rule
(\ref{ZC2}) can be reversed yielding the following transformed
Pauli matrices ($k = 1, 2, 3$).
\begin{eqnarray}
\label{ZC3}
\sigma^\prime_k&=&
\frac{1}{\sqrt{3}}
\left(
\begin{array}{*{2}{c}}
1&\sqrt{2}\ {\rm e}^{-i\varphi_k}\\
\sqrt{2}\ {\rm e}^{i\varphi_k}&-1
\end{array}
\right)\\[0.3cm]
&&\hspace{0.5cm}\varphi_k\ = \ \varphi(k)\ =\ 
\varphi_0\ + \ \frac{2\pi}{3}\ k
\end{eqnarray}
Here, $\varphi_0$ is some arbitrary real constant which, however, has
been shifted with respect to eq.\ (\ref{A4ac}).\\

\setcounter{section}{3}
\section*{Appendix C}
\setcounter{equation}{0}

In the present Appendix we want to illustrate
the formalism used in the main part of the paper in the 
rather trivial case of the real Clifford algebra C(2,1). 
We display the equations (including the notation) 
in close analogy to the discussion 
performed in the main part of the paper. We 
start with some explicit expressions for the gamma
matrices ($\sigma_k$ are the standard Pauli matrices (\ref{ZC1})).

\parbox{7cm}{
\begin{eqnarray}
\label{ZB1a}
\gamma_1&=&\sigma_3\ =\ 
\left(
\begin{array}{*{2}{c}}
1&0\\ 
0&-1
\end{array}
\right)
\end{eqnarray}
}
\hfill
\parbox{7cm}{
\begin{eqnarray}
\label{ZB1b}
\gamma_2&=&\sigma_1\ =\ 
\left(
\begin{array}{*{2}{c}}
0&1\\ 
1&0
\end{array}
\right)
\end{eqnarray}
}

\begin{eqnarray}
\label{ZB1c}
\gamma_3&=&\ i\sigma_2\ =\ 
\left(
\begin{array}{*{2}{c}}
0&1\\ 
-1&0
\end{array}
\right)
\end{eqnarray}

\noindent
The eigenspaces of the gamma matrices $\gamma_1$, $\gamma_2$ 
are described by the following matrices.

\parbox{7cm}{
\begin{eqnarray}
\label{ZB2a}
{\bf A}_{1,1}&=&
\frac{1}{\sqrt{2}}
\left(
\begin{array}{*{1}{c}}
1\\ 
1
\end{array}
\right)
\end{eqnarray}
}
\hfill
\parbox{7cm}{
\begin{eqnarray}
\label{ZB2b}
{\bf A}_{1,-1}&=&
\frac{1}{\sqrt{2}}
\left(
\begin{array}{*{1}{c}}
1\\ 
-1
\end{array}
\right)
\end{eqnarray}
}

\parbox{7cm}{
\begin{eqnarray}
\label{ZB3a}
{\bf A}_{2,1}&=&
\left(
\begin{array}{*{1}{c}}
1\\ 
0
\end{array}
\right)
\end{eqnarray}
}
\hfill
\parbox{7cm}{
\begin{eqnarray}
\label{ZB3b}
{\bf A}_{2,-1}&=&
\left(
\begin{array}{*{1}{c}}
0\\ 
1
\end{array}
\right)
\end{eqnarray}
}

\begin{table}[t]
\unitlength1.mm
\begin{picture}(150,90)
\put(35,5){
\begin{picture}(80,80)
\linethickness{0.15mm}
\put(0,40){\vector(1,0){80}}
\put(40,0){\vector(0,1){80}}
\put(0,0){\line(1,1){80}}
\put(0,80){\line(1,-1){80}}
\put(40,40){\line(5,2){4}}
\put(45,42){\line(5,2){4}}
\put(50,44){\line(5,2){4}}
\put(55,46){\line(5,2){4}}
\put(60,48){\line(5,2){4}}
\put(65,50){\line(5,2){4}}
\put(70,52){\line(5,2){4}}
\put(75,54){\line(5,2){4}}
\put(40,40){\line(-5,-2){4}}
\put(35,38){\line(-5,-2){4}}
\put(30,36){\line(-5,-2){4}}
\put(25,34){\line(-5,-2){4}}
\put(20,32){\line(-5,-2){4}}
\put(15,30){\line(-5,-2){4}}
\put(10,28){\line(-5,-2){4}}
\put(5,26){\line(-5,-2){4}}
\put(40,40){\line(2,-5){1.6}}
\put(42,35){\line(2,-5){1.6}}
\put(44,30){\line(2,-5){1.6}}
\put(46,25){\line(2,-5){1.6}}
\put(48,20){\line(2,-5){1.6}}
\put(50,15){\line(2,-5){1.6}}
\put(52,10){\line(2,-5){1.6}}
\put(54,5){\line(2,-5){1.6}}
\put(40,40){\line(-2,5){1.6}}
\put(38,45){\line(-2,5){1.6}}
\put(36,50){\line(-2,5){1.6}}
\put(34,55){\line(-2,5){1.6}}
\put(32,60){\line(-2,5){1.6}}
\put(30,65){\line(-2,5){1.6}}
\put(28,70){\line(-2,5){1.6}}
\put(26,75){\line(-2,5){1.6}}
\put(75,36){\parbox[b]{4mm}{$x_1$}}
\put(41,76){\parbox[b]{4mm}{$x_2$}}
\put(82,82){\parbox[b]{10mm}{$A_{1,1}$}}
\put(-7,82){\parbox[b]{10mm}{$A_{1,-1}$}}
\put(82,38){\parbox[b]{10mm}{$A_{2,1}$}}
\put(38.5,82){\parbox[b]{10mm}{$A_{2,-1}$}}
\put(82,55){\parbox[b]{25mm}{$A_{1+} = A_{-1-}$}}
\put(13,82){\parbox[b]{25mm}{$A_{1-} = A_{-1+}$}}
\end{picture}  }
\end{picture}

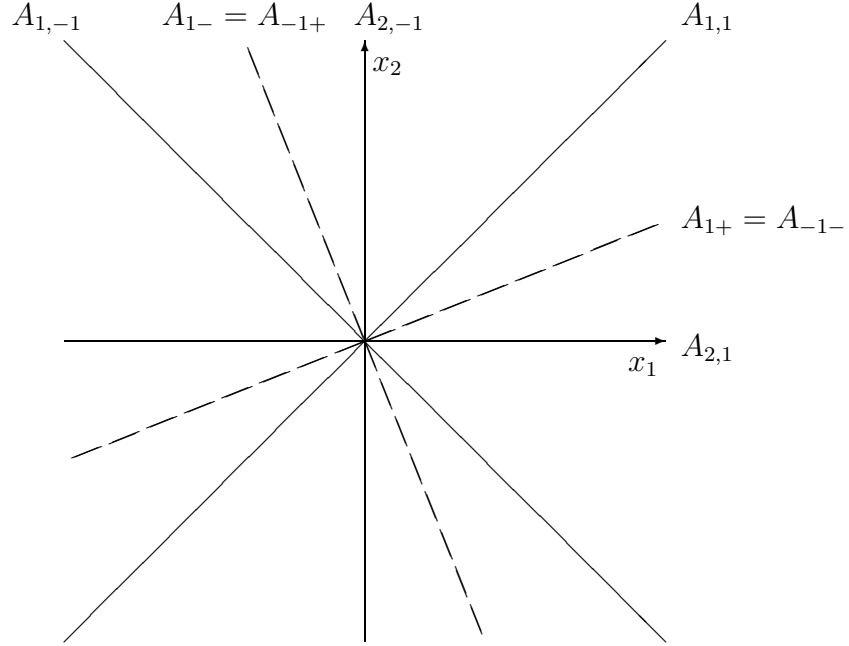
{\bf Figure 1:} Geometry of the eigenspaces of the gamma
matrices $\gamma_1$, $\gamma_2$ [(\ref{ZB1a}), (\ref{ZB1b})]
\end{table}

\noindent
It is clear that the angle between the eigenspaces (lines, 1-planes) 
which relate to different gamma matrices 
$\gamma_1$, $\gamma_2$ is $\pi/4$ (cf.\ Fig.\ 1)\footnote{A related
discussion can be found in \cite{somm}, Vol.\ 2, Mathematische Zus\"atze
und Erg\"anzungen [Mathematical Supplements], no.\ 
13: Zwei- und vierreihige Matrizen. Darstellung der 
hyperkomplexen $\gamma$-Einheiten durch Matrizen. Zu Kap.\ IV, \S 5
[Two- and four-row matrices. Representation of the hypercomplex
$\gamma$ units by matrices. To Ch.\ IV, \S 5],
p.\ 780 of the cited German edition, also earlier editions
of \cite{somm}, Vol.\ 2, contain this material in some appendix.}. 
Each line through the origin ${\bf x} = (0,0)$ is 
(trivially) isoclinic to each other such line. Therefore,
the analogues of eqs.\ (\ref{C3a}), (\ref{C3c}) are
\begin{eqnarray}
\label{ZB4}
x_2&=&\lambda\ x_1\ \ ,\\[0.3cm]
x_1&=&\lambda^\prime\ x_2\ \ ,\ \ \ \lambda^\prime\ =\ \lambda^{-1}\ \ .
\end{eqnarray}
Eqs.\ (\ref{C4a}), (\ref{C4b}) are mirrored by
\begin{eqnarray}
\label{ZB5a}
{\bf B}(\lambda)&=&
\frac{1}{\sqrt{1 +\lambda^2}}
\left(
\begin{array}{*{1}{c}}
1\\
\lambda
\end{array}
\right)\ \ \ ,
\end{eqnarray}
and
\begin{eqnarray}
\label{ZB5b}
{\bf B}(\lambda^\prime)&=&
\frac{1}{\sqrt{1 +\lambda^{\prime\; 2}}}
\left(
\begin{array}{*{1}{c}}
\lambda^\prime\\
1
\end{array}
\right)\ \ \ .
\end{eqnarray}
Of course, to each set of the
eigenspaces $\{A_{1,1},A_{2,1} \}$, $\{A_{1,-1},A_{2,-1} \}$
two lines $A_{1\pm}$, $A_{-1\pm}$ exist, 
respectively, which lie symmetric with respect to the 
elements of the set (cf.\ Fig.\ 1). 
The analogue of eq.\ (\ref{C5}) reads ($\theta_k$ are the 
angles between the two eigenspaces to the eigenvalue 
$\rho = 1$ [$\rho = -1$] and $A_{1\pm}$ [$A_{-1\pm}$])
\begin{eqnarray}
\label{ZB6}
\cos^2 2\theta_1\ +\ \cos^2 2\theta_2&=&1\ \ .
\end{eqnarray}
For $A_{1\pm}$, $A_{-1\pm}$ the relations
(in analogy to eqs.\ (\ref{C6}), (\ref{C7}))
\begin{eqnarray}
\label{ZB7}
\theta_1&=&\theta_2\ =\ \theta_{sym}\ \ ,\ \ \ 
\cos 2\theta_{sym}\ =\ \pm \frac{1}{\sqrt{2}}
\end{eqnarray}
and
\begin{eqnarray}
\label{ZB8}
w&=&\cos^2 \theta_{sym}\ =\ \frac{1}{2}\ \left(1\ +\
\cos 2\theta_{sym}\right)\ =\ 
\frac{1}{2}\ \left(1\ \pm \frac{1}{\sqrt{2}}\right)\ = \ w_\pm
\end{eqnarray}
are valid.\\

In analogy to eqs.\  (\ref{D1a})-(\ref{D2c}), 
in order to determine the lines $A_{1\pm}$, $A_{-1\pm}$ we have to solve
the following equations for $\rho = 1$ (in sequence for the indices $k = 1$, 
$k = 2$ of the gamma matrices, respectively)

\parbox{7cm}{
\begin{eqnarray}
\label{ZB9a}
w_{\pm}&=&\frac{\left(1 + \lambda^\prime\right)^2}{
2\left(1 + \lambda^{\prime\; 2}\right)}\ \ ,
\end{eqnarray}
}
\hfill
\parbox{7cm}{
\begin{eqnarray}
\label{ZB9b}
w_{\pm}&=&\frac{\lambda^{\prime\; 2}}{1 + \lambda^{\prime\; 2}}\ \ ,
\end{eqnarray}
}

\noindent
and for $\rho = -1$,

\parbox{7cm}{
\begin{eqnarray}
\label{ZB10a}
w_\pm&=&\frac{\left(1 - \lambda\right)^2}{
2\left(1 + \lambda^2\right)}\ \ ,
\end{eqnarray}
}
\hfill
\parbox{7cm}{
\begin{eqnarray}
\label{ZB10b}
w_\pm&=&\frac{\lambda^2}{
1 + \lambda^2}\ \ ,
\end{eqnarray}
}

\noindent
(The eqs.\ (\ref{ZB9a}), (\ref{ZB9b}) [(\ref{ZB10a}), (\ref{ZB10b})] have 
been derived using eq.\ (\ref{ZB5b}) [(\ref{ZB5a})].).
The solution of the above equations reads for $\rho = 1$
\begin{eqnarray}
\label{ZB11}
\lambda^\prime&=&\lambda_\pm\ \ ,
\end{eqnarray}
and for $\rho = -1$
\begin{eqnarray}
\label{ZB12}
\lambda&=&-\lambda_\pm\ \ \ .
\end{eqnarray}
Here, 
\begin{eqnarray}
\label{ZB13a}
\lambda_\pm &=& \pm 2\sqrt{2}\ w_\pm
\end{eqnarray}
which entails
\begin{eqnarray}
\label{ZB13b}
\lambda_\pm \lambda_\mp &=& -1\ \ .
\end{eqnarray}
Inserting eqs.\ (\ref{ZB12}) and (\ref{ZB13b}) into the eqs.\ (\ref{ZB5b}) and
(\ref{ZB5a}), respectively, one finds
\begin{eqnarray}
\label{ZB14a}
{\bf A}_{1\pm}&=&
\frac{1}{\sqrt{1 + \lambda^2_{\pm}}}
\left(
\begin{array}{*{1}{c}}
\lambda_\pm\\ 
1
\end{array}
\right)\ \ ,\\[0.3cm]
\label{ZB14b}
{\bf A}_{-1\pm}&=&
\frac{1}{\sqrt{1 + \lambda^2_{\pm}}}
\left(
\begin{array}{*{1}{c}}
1\\
-\lambda_\pm
\end{array}
\right)
\end{eqnarray}
(cf.\ Fig.\ 1; it holds $A_{1+} = A_{-1-}$, $A_{1-} = A_{-1+}$).
The orthogonal transformation leading to the new coordinate
system consequently reads
\begin{eqnarray}
\label{ZB15}
O_\pm&=&
\frac{1}{\sqrt{1 + \lambda^2_{\pm}}}
\left(
\begin{array}{*{2}{c}}
\lambda_\pm&1\\ 
1&-\lambda_\pm
\end{array}
\right)\ \ .
\end{eqnarray}
This way the following final result is obtained.

\parbox{7cm}{
\begin{eqnarray}
\label{ZB16a}
\gamma^\prime_{1\pm}&=&
\pm \frac{1}{\sqrt{2}}
\left(
\begin{array}{*{2}{c}}
1&1\\ 
1&-1
\end{array}
\right)\ \ \ \ \ \
\end{eqnarray}
}
\hfill
\parbox{7cm}{
\begin{eqnarray}
\label{ZB16b}
\gamma^\prime_{2\pm}&=&
\pm \frac{1}{\sqrt{2}}
\left(
\begin{array}{*{2}{c}}
1&-1\\ 
-1&-1
\end{array}
\right)\ \ \ \ \ \
\end{eqnarray}
}

\begin{eqnarray}
\label{ZB16c}
\gamma^\prime_3&=&
\left(
\begin{array}{*{2}{c}}
0&-1\\ 
1&0
\end{array}
\right)
\end{eqnarray}
It is clear that in the present case there is no 
residual continuous symmetry which has been exploited in 
sect.\ 5 of the main part of the paper which is dealing 
with the real Clifford algebra C(3,1).\\

\end{document}